\input{aipcheck}

\documentclass[
    ,final            
  ]
  {aipproc}

\layoutstyle{6x9}
\usepackage{amsmath,amssymb}
\usepackage{wasysym}
\usepackage{graphicx}
\usepackage{sidecap}
\usepackage{subfigure}
\newcommand{\ME}{\mathcal{M}}

\newcommand{\TeV}{{\ensuremath\rm TeV}}

\newcommand{\chp}{\tilde{\chi}^+}
\newcommand{\chm}{\tilde{\chi}^-}

\newcommand{\om}{\texttt{O'Mega}}
\newcommand{\whizard}{\texttt{WHIZARD}}

\newcommand{\GeV}{{\ensuremath\rm GeV}}
\newcommand{\eqn}{equation}

\newcommand{\ab}{{\ensuremath\rm ab}}

\begin{document}
\rightline{DESY 06-189}
\title{NLO Simulations of Chargino Production at the ILC}
\classification{{12.15.Lk}, 
  {13.40.Ks},
  {13.66.Hk},
  {14.80.Ly}}
\keywords      {Supersymmetry, NLO, Monte Carlo Event Generator }

\author{W. Kilian}{
  address={Fachbereich Physik, University of Siegen, D--57068 Siegen, Germany}, ,altaddress={Deutsches Elektronen-Synchrotron DESY, D--22603 Hamburg, Germany}
}

\author{J. Reuter}{
  address={Carleton University, Dep. of Physics, 1125 Colonel By
  Dr., Ottawa, ON, K1S 5B6, Canada}, ,altaddress={Deutsches Elektronen-Synchrotron DESY, D--22603 Hamburg, Germany}
}

\author{\underline{T. Robens}}{
  address={Deutsches Elektronen-Synchrotron DESY, D--22603 Hamburg, Germany}
}

\begin{abstract}
 We present an extension of the Monte Carlo Event Generator \whizard 
 which includes chargino production at the ILC at NLO. We present two ways
 of adding photonic contributions. We present results for cross
 sections and event generation. 
\end{abstract}

\maketitle


\section{Introduction}
In many GUT models, the masses of charginos tend to be near the lower
edge of the superpartner spectrum, and  they can be pair-produced at a first-phase ILC with c.m.\ energy of
$500\;\GeV$. The precise measurement of their parameters (masses,
mixings, and couplings) is a key for uncovering 
the fundamental properties of the
MSSM \cite{Aguilar-Saavedra:2005pw}. Regarding the
experimental precision at the ILC, off-shell kinematics for the signal process, and
the reducible and irreducible backgrounds \cite{Hagiwara:2005wg} need to be included as well as NLO corrections for chargino
production at the ILC which are in the percent regime. We here present the inclusion of the latter.


\section{Chargino production at LO and NLO}


The total fixed-order NLO cross section is given by
\begin{\eqn}\sigma_{\rm tot}(s,m_e^2) = \sigma_{\rm Born}(s) + 
  \sigma_{\rm v+s}(s,\Delta E_\gamma,m_e^2) + 
  \sigma_{\rm 2\rightarrow 3}(s,\Delta E_\gamma,m_e^2),
\end{\eqn}
where $s$ is the cm energy, $m_{e}$ the electron mass, and
$\Delta\,E_{\gamma}$ the soft photon energy cut dividing the photon
phase space.  
The 'virtual' 
contribution is the interference of the one-loop corrections
\cite{Fritzsche:2004nf} with the Born term. The collinear and infrared
singularities are regulated by $m_e$ and $\lambda$, respectively. 
The dependence on $\lambda$ is eliminated by
adding the soft real photon contribution $\sigma_{\rm soft}
\,=\,f_{\rm soft}\,\sigma_{\rm Born}(s)$ with a universal soft factor 
$f_{\rm soft}(\frac{\Delta E_\gamma}{\lambda})$
\cite{Denner:1991kt}. We break the `hard' contribution 
$\sigma_{\rm 2\rightarrow 3}(s,\Delta E_\gamma,m_e^2)$, i.e., the
real-radiation process $e^-e^+\rightarrow\chm_i\chp_j\gamma$,
into a
collinear and a non-collinear part, separated at a photon
acollinearity angle $\Delta\theta_\gamma$ relative to the incoming
electron or positron:
$
  \sigma_\text{$2\to 3$}(s,\Delta E_\gamma,m_e^2) =
  \sigma_\text{hard,non-coll}(s,\Delta E_\gamma,\Delta\theta_\gamma)
  + \sigma_\text{hard,coll}(s,\Delta E_\gamma,\Delta\theta_\gamma,m_e^2).
$
The collinear part is approximated by convoluting the Born cross section with a
structure function $f(x;\Delta\theta_\gamma,\frac{m_e^2}{s})$
\cite{Bohm:1993qx}. The non-collinear part is generated
explicitely.
 
The total fixed order cross section 
is implemented in the multi-purpose event generator
\om/\whizard \cite{Moretti:2001zz,Kilian:2001qz} using 
a `user-defined' structure 
function and an effective matrix
element $|\ME_\text{eff}|^2$ which contains the
Born part, the soft-photon factor and the Born-1 loop interference
term.  
\begin{figure}
\centering
\includegraphics[width=.45\textwidth]{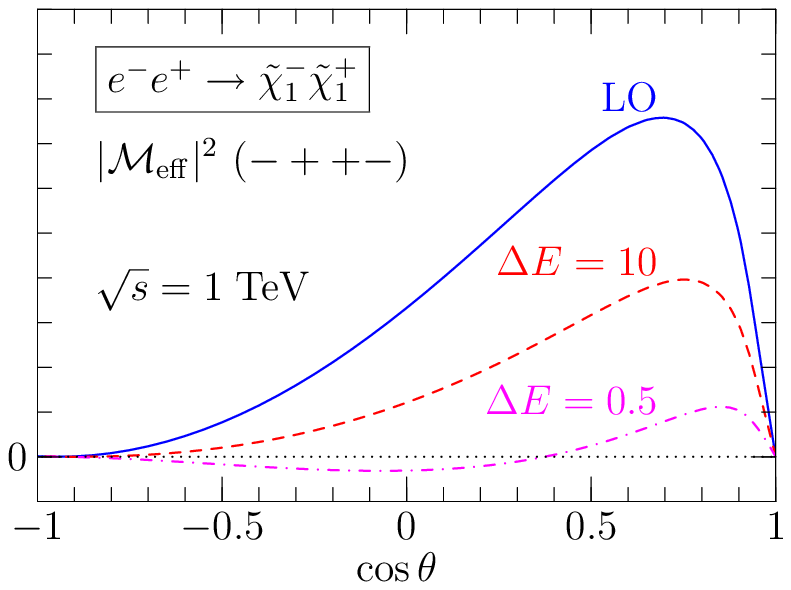} \quad
\includegraphics[width=.45\textwidth]{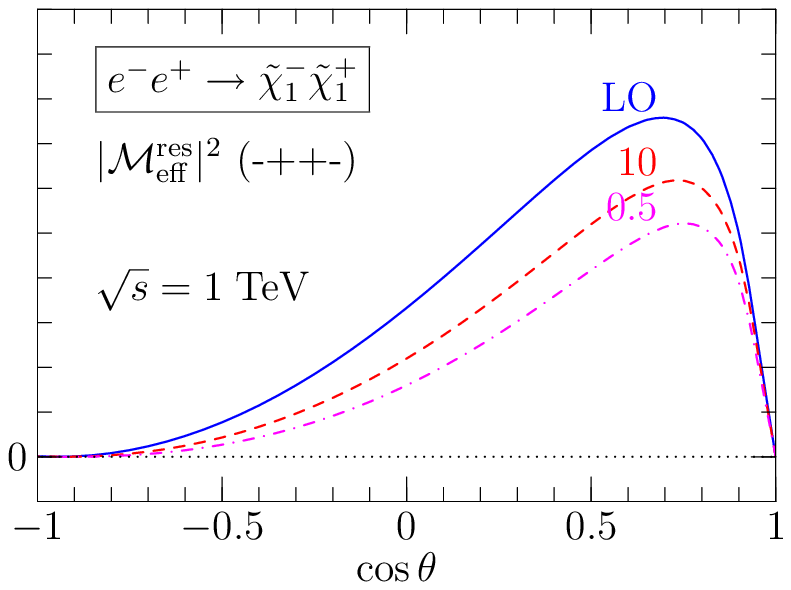}
\caption{$\theta$ -dependence of effective squared matrix element 
  ($\sqrt{s}=1\;\TeV$).Left figure: fixed order effective matrix element; right
  figure: effective matrix element with
  the one-photon ISR part subtracted.  Solid line: Born term; dashed: including virtual
  and soft contributions for $\Delta E_\gamma=10\;\GeV$; dotted: same with
  $\Delta E_\gamma=0.5\;\GeV$.  
  $\Delta\theta_\gamma=1^\circ$.}
\label{fig:Meff}
\end{figure}
In the soft-photon region this approach runs into
the problem of negative event
weights \cite{Bohm:1986fg, Kleiss:1989de}: for some values of $\theta$,  
the $2\to 2$ part of the NLO-corrected
squared matrix element is positive definite by itself only if $\Delta
E_\gamma$ is sufficiently large and becomes negative otherwise, cf Fig. \ref{fig:Meff}.
To still obtain unweighted event samples, an ad-hoc approach is to simply drop events with negative events before
proceeding further.

Alternatively, negative event weights can be eliminated using an
exponentiated structure function 
$f_\text{ISR}$ which resums
higher-order initial
radiation \cite{Gribov:1972rt,Skrzypek:1990qs}. The cancellation of
infrared singularities between virtual and real corrections is built-in, so
that the main source of negative event weights is eliminated. To
combine the ISR-resummed LO result with the additional NLO 
contributions \cite{Fritzsche:2004nf}, we subtract from the effective
squared matrix element, for each 
incoming particle, the contribution of one soft photon that has already been
accounted for in $\sigma_\text{s}$.
This defines $|\widetilde\ME_\text{eff}|^2 \, = \,
|\ME_\text{eff}|^{2}- 2f_\text{soft,ISR}  \,|\ME_\text{Born}|^2$,
which contains the Born term, the virtual and soft collinear 
contribution with the leading-logarithmic part of virtual photons and
soft collinear emission removed, and soft non-collinear radiation of
one photon. Convoluting
this with the resummed ISR structure function,
we obtain a modified $2\to 2$ part of the total cross section which
also includes soft and collinear photonic 
corrections to the Born/one-loop interference.  The complete result
also contains the hard non-collinear $2\to 3$ part.
The resummation
approach eliminates the problem of negative weights
(cf. Fig. \ref{fig:Meff}) such that unweighting of generated events and
realistic simulation at NLO are now possible in all regions of
phase-space. A final improvement is to also convolute the $2\to 3$ part with
the ISR structure function \cite{Kilian:2006cj}.


\section{Results}

\begin{figure}
\centering
  \includegraphics[height=0.39\textwidth,width =0.8\textwidth]{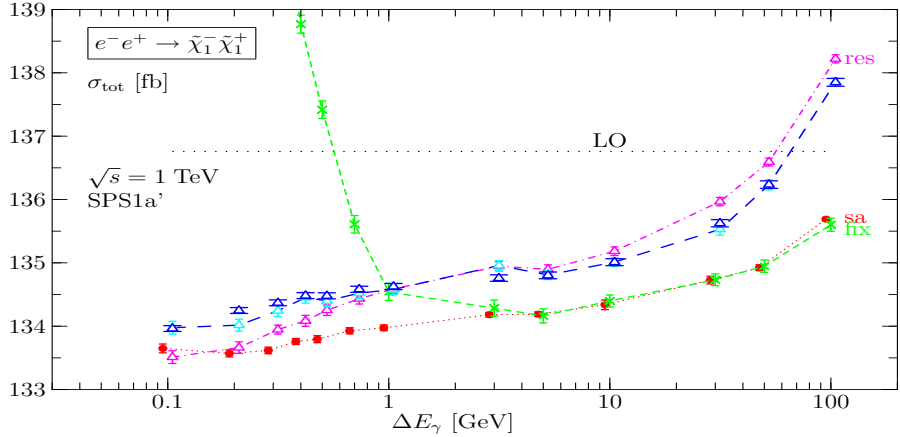}
  \caption{\label{fig:edep} Total cross section dependence on $\Delta
    E_\gamma$: {\rm `sa'}
    (dotted) = fixed-order semianalytic result; {\rm `fix'} (dashed) = fixed-order
    Monte-Carlo result; {\rm `res'}
    (long-dashed) = ISR-resummed Monte-Carlo result; (dash-dotted) = same but resummation applied only to
    the $2\to 2$ part. $\Delta\theta_\gamma=1^\circ$.
    LO: Born cross section.}
\label{fig:edep}
\end{figure}
First, we have
to check the dependence of the total cross section on the cutoffs $\Delta E_\gamma$, $\Delta\theta_\gamma$.
 Fig. \ref{fig:edep} compares the $\Delta E_{\gamma}$ dependence of
 the numerical results from 
the semianalytic fixed-order calculation with the Monte-Carlo
integration in the fixed-order and in the resummation schemes. The fixed-order Monte-Carlo result agrees with the semianalytic
result as long as the cutoff is greater
than a few $\GeV$ but departs from it for smaller cutoff values
because here, in some parts of phase 
space, $|\ME_{\rm eff}|^{2}\,<\,0$ is set to zero. The semianalytic
fixed-order result is not exactly 
cutoff-independent, but exhibits 
a slight rise of the calculated cross section with increasing cutoff
(breakdown of the soft approximation). For $\Delta E_\gamma=1\;\GeV$
($10\;\GeV$) the shift is about 
$2\,\permil$ ($5\,\permil$) of the total cross section. The fully resummed result  shows an increase of about
$5\,\permil$ of the total cross section with respect to the
fixed-order result which stays roughly constant until $\Delta
E_\gamma>10\;\GeV$.  This is due to higher-order photon radiation.

For the dependence on the collinear cutoff $\Delta\theta_\gamma$, the main higher-order effect is associated with
photon emission angles below $0.1^\circ$. For $\theta_\gamma>10^\circ$,
the collinear approximation breaks down. 

\begin{figure}
\centering
  \includegraphics[height=.35\textwidth,width=.95\textwidth]{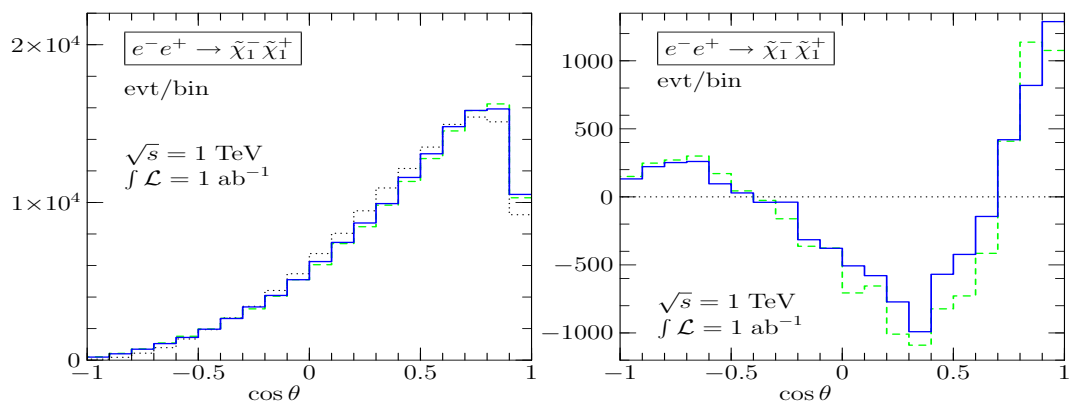}
  \vspace{\baselineskip}
  \caption{Polar scattering angle distribution for an integrated
    luminosity of $1\;\ab^{-1}$ at $\sqrt{s}=1\;\TeV$. Left: total
    number of events per bin; right: difference w.r.t.\ the Born
    distribution.  LO (dotted) = Born cross section without
    ISR; fix (dashed) = fixed-order approach; res (full)
    = resummation approach.  Cutoffs: $\Delta E_\gamma=3\;\GeV$ and
    $\Delta\theta_\gamma=1^\circ$.}
\label{fig:histth}
\end{figure}

In
Fig. \ref{fig:histth} we show the binned distribution of the chargino
production angle as obtained from a sample of unweighted events. 
It demonstrates that NLO corrections (which, for total cross sections, are in the percent regime and can reach $20\%$ at the threshold) are important and cannot be
accounted for by  a constant K factor. In summary, to carefully choose the
resummation method and cutoffs will be critical for a truly precise
analysis of real ILC data. For more details, cf. \cite{Kilian:2006cj,Robens:2006np}.


\section{Conclusions}
We have implemented NLO corrections
into the event generator \whizard for chargino pair-production at the
ILC with several approaches for the inclusion of photon radiation. 
A careful analysis of the dependence on the cuts
$\Delta\,E_{\gamma},\,\Delta\,\theta$  reveals 
uncertainties related to higher-order radiation and breakdown of the
soft or collinear approximations. The version of the program resumming
photons allows to get rid of negative event weights in the simulation,
accounts for all yet known higher-order effects, allows for 
cutoffs small enough that soft- and collinear-approximation artefacts
are negligible, and explicitly generates photons where they can be
resolved experimentally. Corrections for the decays of charginos and
non-factorizing corrections are in the line of future work. 

\begin{theacknowledgments}
This
work was supported by the German Helmholtz Association, Grant
VH--NG--005.
\end{theacknowledgments}



\bibliographystyle{aipproc}   

\bibliography{litmod}

\IfFileExists{\jobname.bbl}{}
 {\typeout{}
  \typeout{******************************************}
  \typeout{** Please run "bibtex \jobname" to optain}
  \typeout{** the bibliography and then re-run LaTeX}
  \typeout{** twice to fix the references!}
  \typeout{******************************************}
  \typeout{}
 }

\end{document}